\documentclass[english,aps,pra,twocolumn,showpacs,supferscriptaddress,floatfix]{revtex4-1}
\usepackage[T1]{fontenc}
\usepackage[latin1]{inputenc}
\usepackage{amsmath}
\usepackage{graphicx}
\usepackage{amssymb}
\usepackage{dcolumn} 
\usepackage{bm} 

\makeatletter

\usepackage{amsfonts}
\usepackage{epsfig}
\usepackage{dsfont}

\newcommand{\Eq}[1]{Eq.~(\ref{#1})}

\newcommand{\be}{\begin{equation}}
\newcommand{\bea}{\begin{eqnarray}}
\newcommand{\eea}{\end{eqnarray}}
\newcommand{\ee}{\end{equation}}

\newcommand{\bra}[1]{\mbox{$\langle #1 |$}}
\newcommand{\ket}[1]{\mbox{$| #1 \rangle$}}

\newcommand{\quot}[1]{``#1''}

\newcommand{\U}{{\cal U}}
\newcommand{\C}{{\cal C}}

\newcommand{\G}{{\cal G}}

\usepackage{babel}
\makeatother

\begin{document}

\title{Genuine fidelity gaps associated with a sequential decomposition of genuinely entangling isometry and unitary operations}


\author{Hamed \surname {Saberi}}
\email{hamed.saberi@lmu.de}
\affiliation{Department of Physics, Shahid Beheshti University, G.C., Evin, Tehran 19839, Iran}
\affiliation{School of Physics, Institute for Research in Fundamental Sciences (IPM), Tehran 19395-5531, Iran}

\date{June 3, 2013}

\begin{abstract}

  We draw attention to the existence of \quot{genuine} fidelity gaps in an ancilla-assisted sequential decomposition of genuinely entangling isometry and unitary operations of quantum computing. The gaps arise upon a bipartite decomposition of a multiqubit operation in a one-way sequential recipe in which an ancillary system interacts locally and only once with each qubit in a row. Given the known \quot{no-go} associated with such a theoretically and experimentally desirable decomposition, various figures of merit are introduced to analyze the optimal
  \quot{fidelity} with which an arbitrary genuinely entangling operation may admit such a sequential decomposition. An efficient variational matrix-product-operator (VMPO) protocol is invoked in order to obtain numerically the minimal values of the fidelity gaps incurred upon sequential decomposition of genuine entanglers. We term the values of the gaps so obtained genuine in the light of possible connections to the concept of the genuine multipartite entanglement and since they are independent of the ancilla dimension and the initial states the associated unitaries act upon.

\end{abstract}

\pacs{03.67.Lx; 03.67.Bg; 03.65.Ta; 02.70.-c; 71.27.+a}
 	

\maketitle

\section{Introduction}
\label{sec:intro}

The reductionist approach in devising clever decompositions for expressing the physics of multipartite quantum systems in terms of the information associated with their constituent partitions prevails in all areas of quantum computation and quantum information~\cite{Nielsen2000}. Paradigmatic examples include the Schmidt decomposition for both states~\cite{Schmidt1906,Ekert1995} and operators~\cite{Nielsen1998,Nielsen2003} and the canonical decomposition of two-qubit unitaries~\cite{Khaneja2001,Kraus2001,Duer2002}. From practical point of view, the top-down theoretical search for such decompositions can provide bottom-up and constructive recipes for experimental realization of the original quantum compositions. The universal set of unitary operations already offers the possibility for decomposition of a generic multiqubit unitary operation in terms of a quantum circuit made out of one- and two-qubit gates~\cite{DiVincenzo1995,Barenco1995}.
However, the decomposition may in general require a highly complicated combination of exponentially many gates to invoke and the same qubits can be touched repeatedly at different stages of the prescribed decomposition~\cite{Nielsen2000}. As such, a blind implementation of such a recipe may face daunting theoretical and experimental challenges. Nevertheless, efficient implementation of such \emph{genuinely entangling} multiqubit operations is essential for the purpose of scalable quantum computation.

\begin{figure}[t]
\centering
\includegraphics[width=1\linewidth]{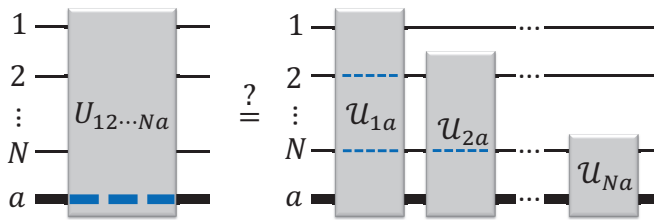}
\caption{Quantum circuit representation of a sequential decomposition of a nonlocal multiqubit unitary $U_{12 \cdots Na}$ into a sequence of
bipartite ancilla-qubit unitaries $\U_{ka}$ that act \emph{only} on the Hilbert space of qubit $k$ and ancilla $a$ while leaving other qubits unchanged.
The blue dashed-line continuation of the qubits' solid lines inside each bipartite unitary emphasizes the trivial action of the ancilla as an identity operator upon such qubits. Ancilla states are also shown with solid bold lines throughout. Note that in order that the desired sequential decomposition can be realized deterministically without the need of performing measurement on the ancilla state, a decoupling ancilla requirement shall be fulfilled on both sides of the graphical equation. The point has been illustrated by the blue dashed-line on the left hand side implying the equivalence $U_{12\cdots Na}= U_{12\cdots N} \otimes \mathds{1}_{a}$.}
\label{fig1:seq_dec}
\end{figure}

In an attempt to facilitate the situation, an ancilla-assisted decomposition of a generic $N$-qubit unitary operation $U_{12\cdots N}$ into a one-way sequence of bipartite ancilla-qubit unitary operations $\U_{ka}$ was proposed by Lamata \emph{et al}~\cite{Lamata2008}. According to such a scenario, each qubit is allowed to interact locally and only once with an itinerant ancillary system (e.g., a trapped multilevel atom coupled to a single mode of an optical cavity in the realm of cavity or circuit QED experiments~\cite{Schoen2005,Schoen2007,Saberi2009}) which intervenes to convey indirectly the nonlocal content of the original global operation throughout the register of qubits (see Fig.~\ref{fig1:seq_dec}). More formally, the \emph{sequential decomposition} of $U_{12 \cdots N a}$ augmented by an ancillary system $a$ seeks a sequence of consecutive bipartite decompositions of the form
\begin{eqnarray}
\label{eq:seq_dec}
\nonumber
U_{1 2 \cdots N a} & = & (\U_{1a} \otimes \mathds{1}_{2} \otimes \cdots \otimes \mathds{1}_N) (\mathds{1}_1 \otimes \U_{2a} \otimes \cdots \otimes \mathds{1}_N) \\
& & \qquad \cdots (\mathds{1}_1 \otimes \mathds{1}_2 \otimes \cdots \otimes \U_{Na})  \; ,
\end{eqnarray}
where the right-hand side can also be described by the shorthand notation
\begin{eqnarray}
\label{eq:seq_dec_short}
U^{\mathrm{seq.}}_{12\cdots Na} \equiv \prod_{k=1}^N \biggl(\bigotimes_{k'<k} \mathds{1}_{k'} \biggr) \otimes \U_{ka} \otimes \biggl( \bigotimes_{k''>k} \mathds{1}_{k''} \biggr)   \; .
\end{eqnarray}
Note that the presence of the identities $\mathds{1}_k$'s remind the point that at step $k$ of the decomposition, the two-body unitary operation $\U_{ka}$ entangles the ancilla $a$ with \emph{only} the $k$'th qubit and leaves the other qubits intact.

A bipartite decomposition of that sort for a generic genuinely entangling unitary, however, was shown to be penalized by a non-unitarily decoupling ancilla in the last step~\cite{Lamata2008}. Various proofs for the impossibility of such a sequential decomposition have since then been offered: A \emph{reductio ad absurdum} proof strategy for such a \quot{no-go} theorem implied the incompatibility of unitarity of the last constituent two-qubit operation and the deterministic nature of the decomposition requiring no measurement to be performed on the ancilla~\cite{Lamata2008}. An alternative proof relying on an operator-Schmidt decomposition~\cite{Nielsen1998,Nielsen2003} of the bipartite unitaries suggested the incompatibility of the three required constrains in the decomposition: (i) sequentiality; (ii) unitarity of the bipartite operators $\U_{ka}$; and (iii) the genuinely entangling feature of all bipartite unitaries not allowing their reduction into tensor products of non-entangling local unitaries~\cite{Saberi2011}. In particular, the latter analysis showed that the sequentiality violates the genuinely entangling requirement of the last unitary giving rise to a Schmidt number of unity for the last unitary.

A different pragmatic perspective on the decomposition, nonetheless, aimed at relaxing the condition of a unitarily decoupling ancilla at the price of
ending up an imperfect (less than unity) \quot{fidelity} as a measure of the quality of the decomposition~\cite{Saberi2011}. Early numerical optimization protocols employing the powerful and flexible variational matrix-product operator (VMPO)~\cite{Zwolak2004,Verstraete2004,McCulloch2007,Crosswhite2008,Verstraete2008,Pirvu2010,Saberi2011} technique focused on quantifying such a fidelity or the degree to which a given global operation can be decomposed with subsequent application of bipartite unitary operations. The method devised an \emph{in situ} local variational optimization protocol to find the optimal bipartite ancilla-qubit operations ${\U}_{ka}$'s that lead to a sequentially decomposed version of the original unitary $U_{12 \cdots N a}$ whose action is not perfectly equivalent to $U_{12 \cdots N a}$ but is rather closest to that in some sense. We point out, in analogy with the paradigm of optimal quantum cloning~\cite{Buzek1996,Gisin1997}, approximate or probabilistic implementation of such a decomposition could yet be essential for various quantum informational tasks and applications~\cite{Saberi2012,Saberi2013}. A Frobenius-norm metric was proposed there as the figure of merit to quantify the fidelity of such a decomposition. The numerical analysis gave imperfect fidelity values independent of the initial states and the ancilla dimension. In this work we propose another figure of merit to benchmark the genuineness of such \quot{gaps} in the fidelities based on a $p$-norm metric criterion and compare the outcome to that of the previously introduced Frobenius-norm metric.

This paper is structured as follows: In Sec.~\ref{sec:MPO_seq_dec} we demonstrate that the sequential decomposition naturally leads to an MPO representation with a bond dimension equal to the ancilla dimension. Alternatively one could use an MPO Ansatz for obtaining the optimal bipartite ancilla-qubit unitaries that lead to the maximal value of the fidelity within the variational optimization prescription. The Frobenius-norm metric and the VMPO methodology for obtaining the values of the fidelity gaps is described in Sec.~\ref{sec:Fro_unitary}. Here we complement our previous achievements by numerically exploring the scaling of the fidelity gaps with the number of qubits $N$. In Sec.~\ref{sec:Fro_isometry} we extend the ideas of the sequential decomposition  to global \emph{isometries} and contrast the issue to that of the \emph{unitaries}. A different figure of merit for quantification of the fidelity gaps based on the $p$-norm metric is introduced in Sec.~\ref{sec:p_norm}. Ultimately, Sec.~\ref{sec:conclusions} contains our conclusions and an assessment of the applicability range of the proposed ideas in various areas of quantum technologies.

\section{Matrix-product operator representation of the sequential decomposition}
\label{sec:MPO_seq_dec}

We first demonstrate explicitly how the sequential decomposition outlined in previous section can be rephrased exactly in terms of the one-dimensional subclass of the important hierarchy of tensor network operators (TNO)~\cite{Verstraete2008} namely MPO~\cite{Zwolak2004,Verstraete2004,McCulloch2007,Verstraete2008,Pirvu2010}. To this end, we express each bipartite unitary $\U_{ka}$ in terms of a complete orthonormal basis
\begin{eqnarray}
\label{eq:U_ka_basis}
\U_{ka}=\sum_{i_k,j_k=0}^1 \sum_{\alpha_k,\beta_k=1}^D \U_{j_k,\beta_k}^{i_k,\alpha_k} |i_k \alpha_k \rangle
\langle j_k \beta_k |  \; ,
\end{eqnarray}
where Roman (Greek) letters denote qubit (ancilla) indices. Plugging the latter expansion for the bipartite unitaries $\U_{ka}$ back into the sequential decomposition Eq.~(\ref{eq:seq_dec_short}) while summing already over the ancilla indices and defining
\begin{eqnarray}
\label{eq:U_[k]}
\sum_{\alpha_k \beta_k} \U_{j_k,\beta_k}^{i_k,\alpha_k} |\alpha_k\rangle \langle\beta_k| \equiv \U_{[k]}^{i_k,j_k}  \; ,
\end{eqnarray}
yields
\begin{eqnarray}
\label{eq:MPO_1}
\nonumber
U^{\mathrm{seq.}}_{12\cdots Na} = \prod_{k=1}^N \sum_{i_k,j_k} \U_{[k]}^{i_k,j_k} \otimes \biggl(\bigotimes_{k'<k} \mathds{1}_{k'} \biggr) \otimes |i_k\rangle \langle j_k|\\  \otimes \biggl( \bigotimes_{k''>k} \mathds{1}_{k''} \biggr)  \; ,
\end{eqnarray}
which can, in turn, be cast into an MPO form given by
\begin{eqnarray}
\label{eq:MPO_2}
\nonumber
U^{\mathrm{seq.}}_{12\cdots Na} = \sum_{i_1,j_1} \cdots \sum_{i_N,j_N} \biggl(\prod_{k=1}^N \U_{[k]}^{i_k,j_k} \biggr) \otimes \biggl(\bigotimes_{k=1}^N  |i_k\rangle \langle j_k| \mathds{1}_k \biggr)  \; , \\
\end{eqnarray}
where the trivial action of the identity $\mathds{1}_k$ on the right hand side has been preserved purposefully to stress the ultimate effect of the tensor products of the identities in \Eq{eq:MPO_1} arising from the sequential nature of the decomposition.


The MPO representation just derived allows local access to the bipartite unitaries $\U_{ka}$ and provide powerful tools for representing such multiqubit unitaries as an efficiently-contractible network of multi-index tensors with the possibility to be optimized numerically by means of a variational algorithm with respect to some cost function. In the following section, we shall elaborate on the numerical benefits of such a tensor-network representation upon considering a Frobenius-norm metric for the calculation of the fidelity gaps associated with an approximate realization of the sequential decomposition.

\section{Sequential decomposition of global operations within the Frobenius-norm metric: The map $N \to N$}
\label{sec:Fro_unitary}

\begin{figure}[t]
\centering
\includegraphics[width=0.65\linewidth]{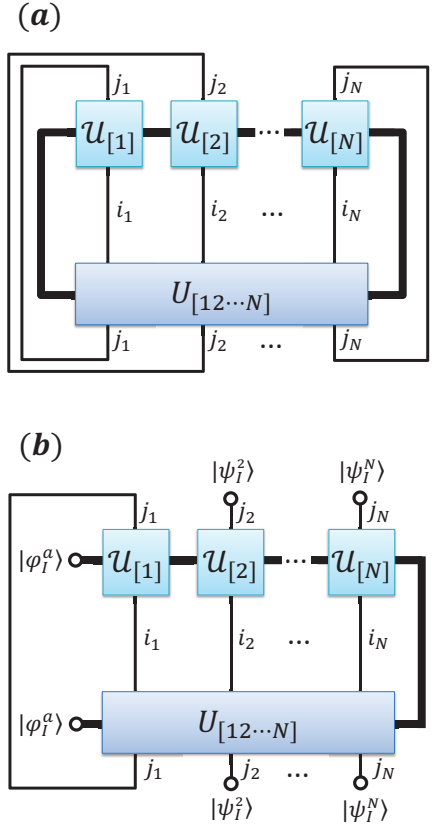}
\caption{Tensor-network representation of Frobenius overlap of MPOs associated with the calculation of the Frobenius cost function \Eq{eq:cost_MPO} for (a) $N \to N$ unitaries and (b) $1\to N$ isometries. The square boxes represent the bipartite ancilla-qubit unitaries $\U_{ka}$ [more precisely the tensors $\U_{[k]}$ in \Eq{eq:U_[k]}], and the rectangular box depicts the target global unitary $U_{12\cdots N}$. The links connecting the boxes also represent either the ancillary or qubit indices that are being contracted (or summed over). Note that a $1\to N$ multiqubit isometry arises as a result of the action of the corresponding unitary $U_{12\cdots N}$ on some initial state of the ancilla denoted by $|\phi_I\rangle$ as well as those of the qubits $|\psi_I\rangle$ save for the first unitary $\U_{[1]}$. Such contraction on initial states has been depicted by open circles in (b).}
\label{fig2:cont_pat}
\end{figure}

In order to examine the precision of the sequential decomposition, one must be able to provide some relevant figure of merit that measures the distance between the \emph{target} global unitary $U_{12 \cdots Na}$ and the one arising from the sequential decomposition $U^{\mathrm{seq.}}_{12\cdots Na}$.
A Frobenius-norm metric~\cite{frob} was already proposed by the present author~\cite{Saberi2011} to quantify the fidelity gap as a trace distance in the Hilbert-Schmidt space of the unitary operators given by
\begin{eqnarray}
\label{eq:cost}
\nonumber
\C^{(F)} &\equiv& \| U_{12 \cdots Na}- U^{\mathrm{seq.}}_{12\cdots Na} \|^2_{F}\\
&=& 2D -2{\rm Re}\lbrace {\rm Tr}[U^\dagger_{12\cdots Na} U^{\mathrm{seq.}}_{12\cdots Na}] \rbrace   \; ,
\end{eqnarray}
and minimize the cost function so defined via an iterative variational optimization procedure within the space of all bipartite unitaries $\U_{ka}$ as the variational parameters. Expressing now the target unitary, too, in terms of the local ancilla and qubits complete orthonormal bases similar to the one done in \Eq{eq:U_ka_basis} for bipartite unitaries
\begin{eqnarray}
\label{eq:Target_sum_anc_2}
\nonumber
U_{12\cdots Na} & = & \sum_{i_1', j_1'} \cdots \sum_{i_N', j_N'} \sum_{\alpha', \beta'} U_{j_1'\cdots j_N', \beta'}^{i_1' \cdots i_N',\alpha'}\\
& & \qquad \times |i_1' \cdots i_N' \alpha'\rangle  \langle j_1' \cdots j_N' \beta'|   \; ,
\end{eqnarray}
and defining accordingly
\begin{eqnarray}
\label{eq:Target_sum_anc_1}
\sum_{\alpha', \beta'} U_{j_1'\cdots j_N', \beta'}^{i_1' \cdots i_N',\alpha'}  \ket{\alpha'} \bra{\beta'} \equiv U_{[12\cdots N]}^{i_1' \cdots i_N',j_1' \cdots j_N'} \; ,
\end{eqnarray}
the Frobenius cost function $\C^{(F)}$ in MPO language boils down to
\begin{eqnarray}
\label{eq:cost_MPO}
\nonumber
\C^{(F)} &=& 2D-2{\rm Re}\biggl\lbrace {\rm Tr}\biggl[\sum_{j'_1, \cdots, j'_N} \sum_{i_1, j_1} \cdots \sum_{i_N, j_N}
{U^{i_1 \cdots i_N, j'_1 \dots j'_N}_{[12\cdots N]}}^\dagger \\
& & \qquad \qquad \biggl(\prod_{k=1}^N \U_{[k]}^{i_k,j_k}\biggr) \otimes \biggl(\bigotimes_{k=1}^N |j'_k\rangle \langle j_k| \biggr) \biggr] \biggr\rbrace \; ,
\end{eqnarray}
which finds a simple and efficiently contractible graphical representation illustrated in Fig.~\ref{fig2:cont_pat}(a). The so-called VMPO approach proposed by the present author then aims at finding the optimal $\U_{[k]}$'s minimizing the Frobenius cost function $\C^{(F)}$ through an iterative local variational optimization technique similar to those already used in the context of matrix-product states (MPS)~\cite{Verstraete2004_3,Verstraete2004_2,Perez2007,Verstraete2008,Saberi2008,Saberi2009,Saberi2009_2,Weichselbaum2009,Weichselbaum2012}) in which one optimizes the bipartite unitaries $\U_{ka}$ one at a time by varying over their matrix elements while keeping the other unitaries fixed, and performs several sweeps through the chain until convergence is reached.

\begin{figure}[t]
\centering
\includegraphics[width=0.7\linewidth]{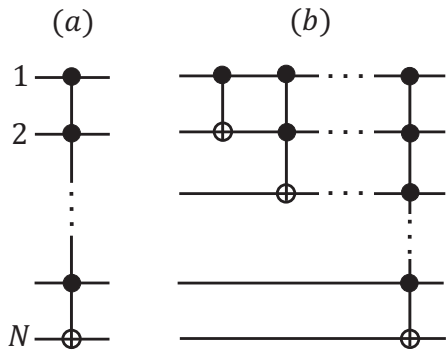}
\caption{Quantum circuit representation of two generalized CNOTs (a) $U_{12\cdots N}^{(1)}$ and (b) $U_{12\cdots N}^{(2)}$ introduced in \Eq{eq:gen_CNOTs} for the sake of analyzing the scaling of the Frobenius fidelity gap with the number of qubits $N$.}
\label{fig3:gen_CNOTs}
\end{figure}

The bipartite ancilla-qubit unitaries used in the variational optimization can in general be expanded in terms of the complete basis of the generators of SU(2) $\otimes$ SU($D$) of the form
\begin{eqnarray}
\label{eq:SU_D}
\U_{[k]}=\exp\biggl(-i\sum_{l=0}^3 \sum_{l'=0}^{2^D-1} h_{l,l'}^{[k]} \sigma_{l} \otimes \tau_{l'} \biggr)   \; ,
\end{eqnarray}
where the real-valued coefficients $h_{l,l'}^{[k]}$ are the variational parameters, $\sigma_1$, $\sigma_2$, $\sigma_3$ denote the usual Pauli sigma matrices with $\sigma_0\equiv \mathds{1}$ the identity matrix, and $\tau_{l'}$ are the generators of SU($D$).

Once the convergence is reached and the minimized value of the cost function $\C^{(F)}$ has been obtained, we may define the normalized Frobenius \emph{fidelity gap} of the sequential decomposition as
\begin{eqnarray}
\label{eq:Frob_Fidelity_gap}
{\G}^{(F)} \equiv {\C^{(F)} \over \|U_{12\cdots Na}\|_F^2+ \|{U_{12\cdots Na}^{\mathrm{seq.}}}\|_F^2}  \; ,
\end{eqnarray}
where the division by the norms in the denominator ensures a value of the gap between zero and one. The results of the application of the outlined procedure to
some paradigmatic gates of quantum computing were reported by the author already in Ref.~\cite{Saberi2011} and for the sake of comparison to the $p$-norm measure to be introduced in the subsequent section can be found again in the first row of Table~\ref{table:fid_gaps}.

Since no change was observed upon increasing the ancilla dimension to the maximal possible values $D=4$, the gaps may be identified as \emph{genuine} of the gates irrespective of the ancilla dimension and the initial state they act upon. The relation to the entangling capabilities and nonlocal character of the target unitary~\cite{Musz2013} with an \emph{operator entanglement} figure of merit~\cite{Zanardi2001,Wang2002,Wang2003} quantified by the \emph{Schmidt strength}~\cite{Nielsen2003} was also discussed by the present author in Ref.~\cite{Saberi2011}.

\begin{figure}[tb]
\centering
\includegraphics[width=1\linewidth]{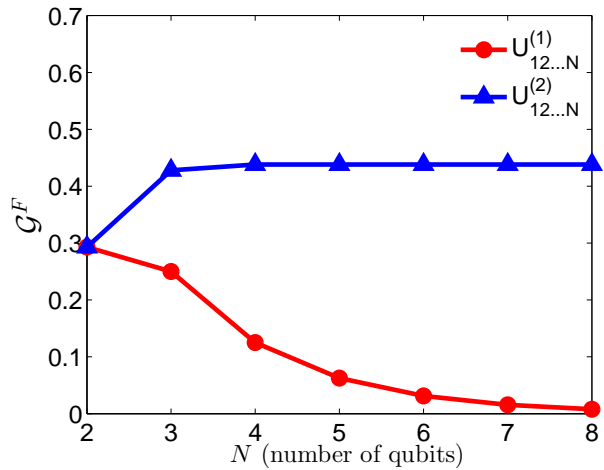}
\caption{Scaling of the Frobenius fidelity gap $\G^{(F)}$ defined in \Eq{eq:Frob_Fidelity_gap} with the number of qubits $N$ for two paradigmatic generalized CNOTs depicted already in Fig.~\ref{fig3:gen_CNOTs}.}
\label{fig4:gaps}
\end{figure}

As a complementary analysis to the previous achievements, we now extend it to the cases beyond hitherto considered three-qubit gates by considering the scaling of the Frobenius fidelity with the number of qubits $N$ for two illustrative generalized CNOTs of the form
\begin{subequations}
\label{eq:gen_CNOTs}
\begin{eqnarray}
U^{(1)}_{12\cdots N} &\equiv& \text{C}^{(N-1)}\text{-NOT}, \\
U^{(2)}_{12\cdots N} &\equiv& \prod_{k=1}^{N} \text{C}^{(k)}\text{-NOT} \otimes \biggl(\bigotimes_{k'=k+2}^{N-1} \mathds{1}_{k'}\biggr) \; ,
\end{eqnarray}
\end{subequations}
depicted schematically in Fig.~\ref{fig3:gen_CNOTs}. Figure~\ref{fig4:gaps} illustrates the results of the application of the VMPO optimization for the calculation of the fidelity gaps for such gates up to $N=8$ qubits and with ancilla of dimension $D=2$. For $U^{(1)}_{12\cdots N}$ an asymptotically vanishing value of the gap is observed upon increasing the number of qubits. Such a behavior might be understood in terms of the fact that $U^{(1)}_{12 \cdots N}$ tends to an identity operator of dimension $2^N\times2^N$ as $N$ increases. On the contrary, for $U^{(2)}_{12\cdots N}$ the gap saturates to a fixed value for large $N$. We have investigated numerically that the matrix representation of a multiqubit gate of the form $U^{(2)}_{12\cdots N}$ in computational basis remains almost the same for all values of $N>2$, which explains the observed behavior.

\section{Sequential decomposition of global isometries within the Frobenius-norm metric: The map $M \to N$}
\label{sec:Fro_isometry}

Here we consider the possibility of sequential decomposition of a global isometry, i.e. the map $M \to N$ when
$M<N$ and calculate the fidelity gaps within the numerical framework described in the last section. Two distinct cases are then identifiable:

(i) \emph{The map $1 \to N$}: For this, one aims at a sequential decomposition of a global isometry with $M=1$, which unlike the $N \to N$ map associated with the genuinely entangling $N$-qubit unitaries was proved to be always possible if the ancilla dimension is large enough~\cite{Lamata2008}. In this case, the corresponding index of only the first
qubit for both the target and the sequentially decomposed operations is free and the rest are contracted with some initial state of the qubits $|\psi_I\rangle$. Here an initial state of the ancilla $|\phi_I\rangle$ shall be also considered. The modified overlap pattern is depicted in Fig.~\ref{fig2:cont_pat} (b).
We have applied our numerics for the case $1 \to 3$ to Toffoli and Fredkin gates and the gap vanished even with
$D=2$. However, we have also considered some randomly-generated $1 \to 3$ isometries and have realized that a nonvanishing fidelity gap
arises with $D=2$ and this is not, though, in contradiction with the fact that an $1 \to N$ map
can always be decomposed sequentially with perfect fidelity, since it could yet require higher ancilla dimension. When we tried out $D=4$ for these gates,
the gap disappeared, as expected.

(ii) \emph{The map $M \to N$ when $1<M<N$}: This may or may not be decomposed sequentially, as demonstrated in Ref.~\cite{Lamata2008}. The numerical framework remains the same as the case of $1 \to N$ except that one would contract on some initial states of the qubits for the last
$N-M$ qubit indices on both the target and the sequential isometries.
Our numerics imply that, for example, the map $2 \to 3$ for Toffoli incurs no gap with $D=4$, whereas randomly generated $2 \to 3$ isometries lead to non-vanishing gaps even with the maximal possible ancilla dimension, i.e., $D=4$, in complete agreement with the theoretical claims of Ref.~\cite{Lamata2008}.

\section{Sequential decomposition of global unitaries within the $p$-norm metric}
\label{sec:p_norm}

In this section we introduce another figure of merit for quantification of the fidelity gaps. The aim is to study the sequential decomposition of unitaries in terms of the $p$-norm of operators instead of the Frobenius one. The same numerical framework developed for the Frobenius
metric applies to the present context, too, except that the Frobenius-norm metric shall be replaced by a $p$-norm ($p=2$) in the corresponding cost function of Eq.~(\ref{eq:cost}) to define a $p$-norm cost function of the form
\begin{eqnarray}
\label{eq:cost_p}
\C^{(p)}=\| U_{12\cdots Na}-{U_{12\cdots Na}^{\mathrm{seq.}}} \|^2_{2}  \; .
\end{eqnarray}

The $p$-norm of a matrix $A$ is defined by~\cite{Golub1996}
\begin{eqnarray}
\label{eq:p_norm}
\|A\|_p=\max_{x \ne 0} {\|A x \|_p \over \|x\|_p}  \; ,
\end{eqnarray}
and in the specific case that $p=2$, it finds the characterization $\|A\|_2=\rho (A^{\dagger} A)^{1/2}$, where $\rho$ denotes the spectral
radius~\footnote{The spectral radius of a matrix $A$ is defined by $\rho\equiv\max_i (|\lambda_i|)$
where $\lambda_i$'s are the eigenvalues of $A$.}. Alternatively, it is known that $\|A\|_2=\max_i (\sigma_i)$ where $\sigma_i$'s are the singular values of $A$~\cite{Golub1996}.

For the normalization of the $p$-norm fidelity gap, we make use of the triangular inequality valid for the $p$-norm according to which~\cite{Golub1996}
\begin{eqnarray}
\label{eq:triangular_ineq}
\|A+B\|_2 \le \|A\|_2+\|B\|_2  \; .
\end{eqnarray}

As a result, the induced normalized $p$-norm fidelity gap may be defined through
\begin{eqnarray}
\label{eq:p_norm_gap}
\G^{(p)} \equiv {{\C}^{(p)} \over (\|U_{12\cdots Na}\|_2+\|{U_{12\cdots Na}^{\mathrm{seq.}}} \|_2)^2}  \; ,
\end{eqnarray}
and is thus guaranteed to change between zero and 1.

On the other hand, it is known that the 2-norm of any unitary matrix is 1. So the normalization factor in this case is always 4,
irrespective of the dimension of the unitaries. Note that in the Frobenius norm case, the normalization factor used to grow with the dimension of the matrices, whereas in the 2-norm case it remains constant.

It is clear from the definition (\ref{eq:p_norm}) that computing $p$-norm is a nonlinear optimization problem over $\mathds{C}^N$ with a non-convex objective
function and the possibility of encountering local extrema~\cite{Higham1992}.
It is a well-known issue in numerical analysis that 2-norm computation is iterative and decidedly
more complicated than the computation of Frobenius norm, 1-norm or $\infty$-norm~\cite{Golub1996,Higham1992}. In the current context, the complexity yet grows with increasing the number of qubits.

The second row of Table~\ref{table:fid_gaps} illustrates the results of the variational minimization of the $p$-norm cost function for obtaining the $p$-norm fidelity gap $\G^{(p)}$ associated with several paradigmatic two- and three-qubit gates and with the maximal ancilla dimension $D=4$. Owing to the above-mentioned numerical instability associated with the calculation of $p$-norm and to achieve reliable precision, an \emph{average} value from a multitude of computer runs has been reported here. Quite interestingly, the $p$-norm gaps follow the same trend as the previously studied Frobenius one: The two locally equivalent gates CNOT and CZ give the same gap, as expected. The smallest and largest gap among two-qubit gates again corresponds to CPHASE and SWAP, respectively. Furthermore, larger values of the gaps compared to the Frobenius measure appear here for three-qubit gates Toffoli and Fredkin which might be rooted in the constant normalization factor employed in the case of the $p$-norm metric in contrast to the increasing one in the Frobenius norm measure.
\begin{table}[t]
\caption{The values of the fidelity gaps for various paradigmatic two- and three-qubit gates according to the Frobenius norm $\G^{(F)}$, $p$-norm $\G^{(p)}$, and the renormalized Frobenius norm metric $\tilde{\G}^{(F)}$ with ancilla of dimension $D=4$.}
\centering
\vspace{5mm}
\begin{tabular}{c c c c c c c }
\hline\hline
$U_{12\cdots N}$: &CNOT &CZ &CPHASE & SWAP & Toffoli & Fredkin  \\ 
\hline
$\G^{(F)}$: & 0.2929 &  0.2929 & 0.0761 & 0.50 & 0.25 & 0.25\\  [.5ex]
$\G^{(p)}$: &0.1480& 0.1480& 0.045& 0.5001& 0.4512& 0.5125\\  [.5ex]
${\tilde{\G}}^{(F)}$: & 0.1464 &  0.1464 & 0.0381 & 0.25 & 0.125 & 0.125\\  [.5ex]
\hline
\hline
\end{tabular}
\label{table:fid_gaps}
\end{table}
Motivated by such a behavior, the new definition for the \quot{renormalized} Frobenius-norm fidelity gap denoted by $\tilde{\G}^{(F)}$ takes the form
\begin{eqnarray}
\label{eq:ren_Frob_gap}
\tilde{{\G}}^{(F)} \equiv {\C^{(F)} \over (\|U_{12\cdots Na}\|_F+\|{U_{12\cdots Na}^{\mathrm{seq.}}} \|_F)^2} \; ,
\end{eqnarray}
with the results in the last row of Table~\ref{table:fid_gaps}. With such normalization scheme,
the $p$-norm gaps remain smaller than the Frobenius one, in complete agreement with the
inequality $\|A\|_2 \le \|A\|_F$ already known from linear algebra~\cite{Golub1996}.

\section{Conclusions and outlook to further research}
\label{sec:conclusions}

In conclusion, we have drawn attention to the emergence of \quot{genuine} fidelity gaps upon a sequential decomposition of genuinely entangling unitary and isometry operations. A flexible theoretical and numerical framework based on the efficient VMPO formalism for quantifying the values of the gaps within the Frobenius and $p$-norm metric has been introduced. The values of the fidelity gaps may be utilized for characterizing the entangling (and disentangling) properties of multiqubit unitary operations. Unveiling possible connections between the genuine fidelity gaps so obtained and the concept of genuine multipartite entanglement might shed light on the important issue of characterization of the genuine multipartite entanglement. We point out,
despite the \quot{no-go} associated with a \emph{deterministic} implementation of the sequential decomposition, possible \emph{probabilistic} implementation of such a decomposition remains yet an open issue.

An exciting application of the sequential decomposition might be realized in the context of quantum key distribution (QKD) protocols utilizing the associated \quot{no-go} for developing effective protection schemes against possible public key disclosure threats in the light of the well-accepted Bennett-Brassard 1984 (BB84) quantum cryptography protocol~\cite{Bennett1984} employing another \quot{no-go} forbidding the cloning of quantum information~\cite{Wootters1982}.


\begin{acknowledgments}

I gratefully acknowledge stimulating discussions with Enrique Solano, Lucas Lamata, David P\'erez-Garc\'{\i}a, Andreas Weichselbaum,
Jan von Delft, and Jens Eisert.
I am also grateful to Universidad del Pa\'{\i}s Vasco for support and hospitality.

\end{acknowledgments}

\bibliography{p_norm}

\begin{thebibliography}{42}
\bibitem [{fro()}]{frob}%
  \BibitemOpen
  \href@noop {} {}\bibinfo {note} {The Frobenius or Hilbert-Schmidt norm of an
  operator is defined by ${\|O\|}_F=\sqrt{\mathrm{Tr}(O^{\dagger} \:
  O)}$.}\BibitemShut {Stop}%
\end{thebibliography}

\end{document}